\documentclass{article}
\usepackage{spconf}
\usepackage{amsmath,amsthm,amsfonts,amssymb}
\usepackage{mathtools}
\usepackage[english]{babel}
\usepackage{float}
\usepackage[pdftex]{graphicx}
\usepackage{color}
\usepackage{multirow}
\usepackage{graphicx}
\usepackage{IEEEtrantools}
\usepackage{bm}
\usepackage{balance}

\usepackage{url}
\usepackage{cite}
\usepackage{comment}

\usepackage[switch]{lineno}
\usepackage[named]{algo}
\usepackage[noend]{algpseudocode}
\usepackage{algorithmicx}
\usepackage{algorithm}
\usepackage{algpseudocode}
\algnewcommand\algorithmicinput{\textbf{INPUTS:}}
\algnewcommand\INPUTS{\item[{\textbf{STAGE 1: COMPUTING INITIAL CONDITIONS}}]}
\algnewcommand\STAGEA{\item[{\textbf{STAGE 2: LINEAR PROGRAM}}]}
\algnewcommand\STAGEB{\item[{\textbf{STAGE 3: CONJUGATE GRADIENT ITERATIONS}}]}
\usepackage{array,multirow}
\usepackage{nicefrac}
\usepackage{mathtools}
\usepackage{multicol}

\usepackage[font=scriptsize,caption=false,labelsep=space]{subfig}

\ninept

\title{K-Space Beamforming for an Array of Quantum Sensors}
\name{Peter Vouras
\thanks{The author acknowledges the many insights into Rydberg quantum sensing gained after discussions with Dr. Alexandra Artusio-Glimpse at the National Institute of Standards and Technology (NIST) in Boulder, Colorado.}
\address{United States Department of Defense, Washington, DC, 20375 USA}
}

\begin{document}
\setlength{\abovedisplayskip}{3pt}
\setlength{\belowdisplayskip}{3pt}

\maketitle

\begin{abstract}
In this paper we present a novel beamforming technique that can be used with an array of quantum sensors.  The transmit waveform is a short-duration frequency comb constructed using a finite number of sinusoidal tones separated by a fixed offset.  Each element in the array is tuned to one of the tones.  When the radiated signal is received by the aperture, each array element accumulates phase at a different rate since it is matched to only one frequency component of the comb waveform.  The result is that over the duration of the received pulse, progressively higher spatial frequencies are generated across the aperture.  By summing the outputs of all the array elements, a strong peak is created in k-space at the precise time instant when the phases of all the array elements align.  The k-space coordinates of the output can then be transformed to angles as discussed in the paper.  This paper also describes how to set waveform parameters and the separation between array elements.  A desirable advantage of the proposed approach is that the received signal is amplified by the coherent integration gain of the entire spatial aperture.
\end{abstract}

\vspace{-8pt}

\section{Emerging Quantum Sensing Technologies}
One of the most intriguing emerging technologies for sensing propagating radio frequency (RF) radiation is the use of Rydberg atom probes. These quantum probes have many unique features that set them apart from traditional antennas and offer several advantages.  First, the measured electric field strength is traceable to the International System of Units (SI)~\cite{gordon2010quantum,gordon2014millimeter,sedlacek2012microwave,holloway2019real,holloway2021multiple,anderson2021self,anderson2022optical,anderson2022atom}.  Furthermore, direct down-conversion of the RF field to baseband by the atoms reduces the need for back-end electronics~\cite{holloway2014broadband}.  Also, intrinsic ultra-wideband tunability from kilohertz to terahertz frequencies with a single probe and a tunable laser is possible~\cite{artusio2022modern,meyer2020assessment,meyer2021waveguide}. 

\section{Conventional Synthetic Aperture and Phased Array Beamforming}
The field ${v(\mathbf{x}, t)}$ radiating outward from a signal source as a function of position ${\mathbf{x}}$ and time ${t}$ is given by,
\begin{equation}
\label{E:eqn01}
    v(\mathbf{x}, t) = e^{-\textrm{j}\frac{2{\pi}}{\lambda}d(\mathbf{x})}e^{\textrm{j}(2{\pi}ft + \phi(t))}
\end{equation}
where ${d(\mathbf{x})}$ is equal to the distance between the signal source and the receive location ${\mathbf{x} = [x \quad y \quad z]^{T}}$, ${f}$ is temporal frequency, and ${\phi(t)}$ is a time-varying phase term due to waveform modulation or Doppler shift.  Eqn. (\ref{E:eqn01}) is valid for any range.  

Beyond the nominal distance of ${2D^2/{\lambda}}$ a plane wave approximation is used to characterize the propagating fields.  Here ${D}$ refers to the largest dimension of the physical antenna and ${\lambda}$ is the operating wavelength.  At far distances, the field of a propagating monochromatic plane wave is approximated by 
\begin{equation}
\label{E:eqn02}
    s(\mathbf{x}, t) = e^{\textrm{j}2{\pi}(-\mathbf{k}^{T}\mathbf{x} + ft) + \textrm{j}\phi(t)}
\end{equation}
where
\begin{align}
\label{E:eqn03}
    \mathbf{k} &= (1/{\lambda})[\sin\theta \cos\phi \quad \sin\theta \sin\phi \quad \cos\theta ]^T \\ \nonumber
    &\triangleq [k_x \quad k_y \quad k_z]^{T}
\end{align}
is the spatial frequency vector.  The angles ${(\theta,\phi)}$ are spherical coordinates and the vector ${\mathbf{k}}$ represents the number of wavelengths per unit distance in each of the three orthogonal spatial directions.

In narrowband phased array or synthetic aperture applications, beamforming is used to focus the array gain towards different directions and to identify signal sources.  The baseband signal outputs ${s_{mn}(t)}$ at time ${t}$ for an array with ${MN}$ elements can be stacked into a spatial vector ${\mathbf{s}(t)}$ as
\begin{equation}
    \mathbf{s}(t) = \left[ \begin{array}{cccc} s_{00}(t) & s_{01}(t) & \ldots & s_{MN-1}(t) \end{array} \right].
\end{equation}
The ${MN \times 1}$ steering vector ${\mathbf{a}(u,v)}$ contains the interelement phase shifts for a narrowband plane wave traversing across the aperture from a direction ${(\theta,\phi)}$,
\begin{align}
\label{E:eqn04}
&\mathbf{a}(u,v) = \\ \nonumber
&\left[ \left. e^{-j\frac{2{\pi}}{\lambda}(md_{x}u_{k} + nd_{y}v_{k})} \right| 0{\leq}m{\leq}M-1, 0{\leq}n{\leq}N-1 \right]^{T}.
\end{align}
where ${m=0,\ldots,M-1}$ is the element index in the x-direction, ${n=0,\ldots,N-1}$ is the element index in the y-direction, ${d_{x}}$ is the spacing between elements in the x-direction, ${d_{y}}$ is the spacing between elements in the y-direction, and
\begin{align}
    u &= \sin\theta\cos\phi  \notag \\
    v &=\sin\theta\sin\phi.
\end{align}
The beamformed output ${b(u,v)}$ in the direction ${(u,v)}$ is formed by taking the dot product of the steering vector ${\mathbf{a}(u,v)}$ and the array output vector ${\mathbf{s}(t)}$ 
\begin{equation}
    b(u,v;t) = \mathbf{a}(u,v)^{H}\mathbf{s}(t).
\end{equation}
The coherent summation of all array element outputs yields an integration gain at the output of the beamformer that increases signal-to-noise-ratio (SNR) by a factor of ${MN}$.

The beamforming operation is equivalent to a spatial Fourier transform and for the planar array of ${M \times N}$ homogeneous elements arranged in the ${xy}$ plane can be written as,
\begin{equation}
\label{E:eqn08}
{b}(u,v;t)= E(u, v) \sum_{m=0}^{M-1}\sum_{n=0}^{N-1}s_{mn}(t)e^{\mathrm{j}k(md_{x}u + nd_{y}v)}.
\end{equation}
Here ${E(u,v)}$ is the element pattern, the wavenumber ${k = 2{\pi}/{\lambda}}$ (note some definitions omit the ${2\pi}$ factor), ${\lambda}$ is the operating wavelength, and ${s_{mn}(t)}$ is the baseband signal sample at the ${mn}$th array element for time instant ${t}$. If the array elements are uniformly spaced on a rectangular grid then the element locations are given by ${x_{m} = md_{x}}$ and ${y_{n} = nd_{y}}$ where ${d_{x}}$ and ${d_{y}}$ denote the distance between elements in the ${x}$ and ${y}$ directions.  

\begin{figure}[]
  \centering
  \includegraphics[width=8cm]{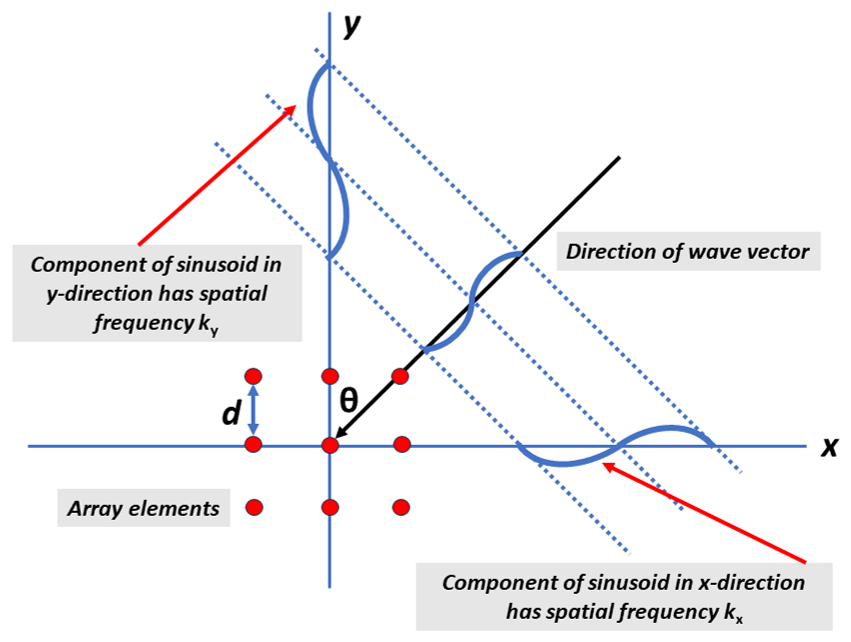}
  \caption{Spatial frequency components of wave propagating in the ${xy}$ plane at an azimuth angle ${\theta}$.  The spacing between array elements is ${d}$.}
  \label{fig:fig01}
\end{figure}

\begin{figure}[]
  \centering
  \includegraphics[width=8cm]{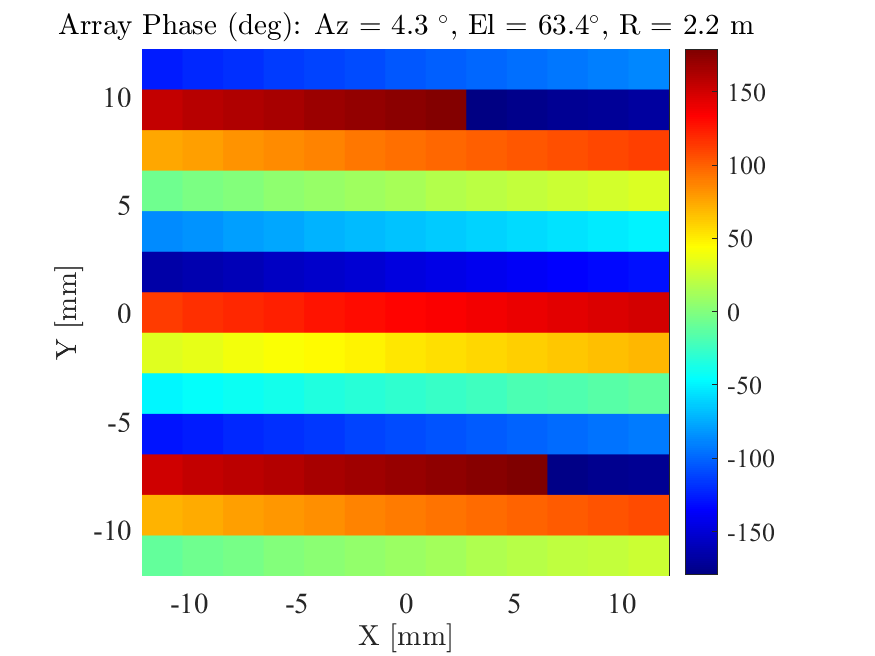}
  \caption{Phase across a planar array for a signal source at a distance of 2.2 meters with an azimuth angle of ${4.3^{\circ}}$ and an elevation angle of ${63.4^{\circ}}$.  Spatial frequency is high in the vertical direction.}
  \label{fig:fig02}
\end{figure}

\begin{figure}[t]
  \centering
  \includegraphics[width=8cm]{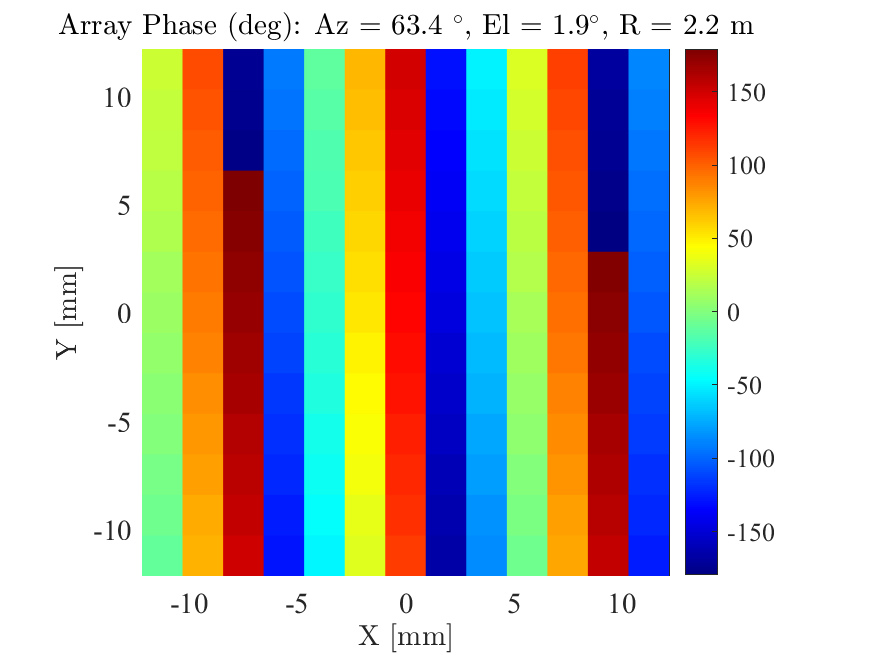}
   \caption{Phase across a planar array for a signal source at a distance of 2.2 meters with an elevation angle of ${1.9^{\circ}}$ and an azimuth angle of ${63.4^{\circ}}$.  Spatial frequency is high in the horizontal direction.}
  \label{fig:fig03}
\end{figure}

The basis functions of the 2-D Fourier transform in (\ref{E:eqn08}) have spatial frequencies given by the wavevector ${\mathbf{k}}$ in (\ref{E:eqn03}).  The spatial frequencies ${k_{x}}$ and ${k_{y}}$ are shown graphically in Fig. \ref{fig:fig01} for a planar array.
The spatial frequency components of an impinging plane wave are revealed in the phase progression across the array.  For a 14-by-14 array with ${\lambda/4}$ spacing at 40 GHz, Fig. \ref{fig:fig02} illustrates the phase (in degrees) at each array element calculated using (\ref{E:eqn01}).  The signal source is at a distance of 2.2 meters with an azimuth angle of ${4.3^{\circ}}$ and an elevation angle of ${63.4^{\circ}}$.  Angles close to array boresight correspond to low spatial frequencies and angles close to ${90^{\circ}}$ yield higher spatial frequencies.  In this case, the signal source is at a large elevation angle so the spatial frequency is high in the vertical orientation.  Fig. \ref{fig:fig03} illustrates the reversed situation where the signal source is at an elevation angle of ${1.9^{\circ}}$ and an azimuth angle of ${63.4^{\circ}}$.  Now the large azimuth angle yields a high spatial frequency in the horizontal direction.

\section{Concept of Operations for K-Space Beamforming}
The transmit signal for the proposed array architecture consists of a uniformly-weighted frequency comb ${s_{t}(t)}$ with finite duration ${T}$ that is the sum of ${N}$ distinct sinusoids separated in frequency by ${\Delta{f}}$ Hz,
\begin{equation}
    \label{E:eqn09}
    s_{t}(t) = \sum_{n=1}^{N}\cos(2{\pi}(f_{0} + n{\Delta}f)t).
\end{equation}
Here ${f_{0}}$ corresponds to the carrier frequency.  Starting from the aperture edge, each array element is tuned to a single, progressively higher frequency of the comb.  After summing together all the element outputs, the signal at the output of the array for a single source is equal to
\begin{equation}
    \label{E:eqn09}
    s_{r}(t) = A\sum_{n=1}^{N}\cos(2{\pi}(f_{0} + n{\Delta}f)t + 2{\pi}d_{n}/{\lambda_n})
\end{equation}
where ${d_{n}}$ corresponds to the distance between the signal source and the ${n}$th array element, ${\lambda_{n}}$ is the wavelength of the ${n}$th frequency, and ${A}$ is the signal amplitude assumed equal for all the sinusoidal tones.  

\begin{figure}[]
  \centering
  \includegraphics[width=8cm]{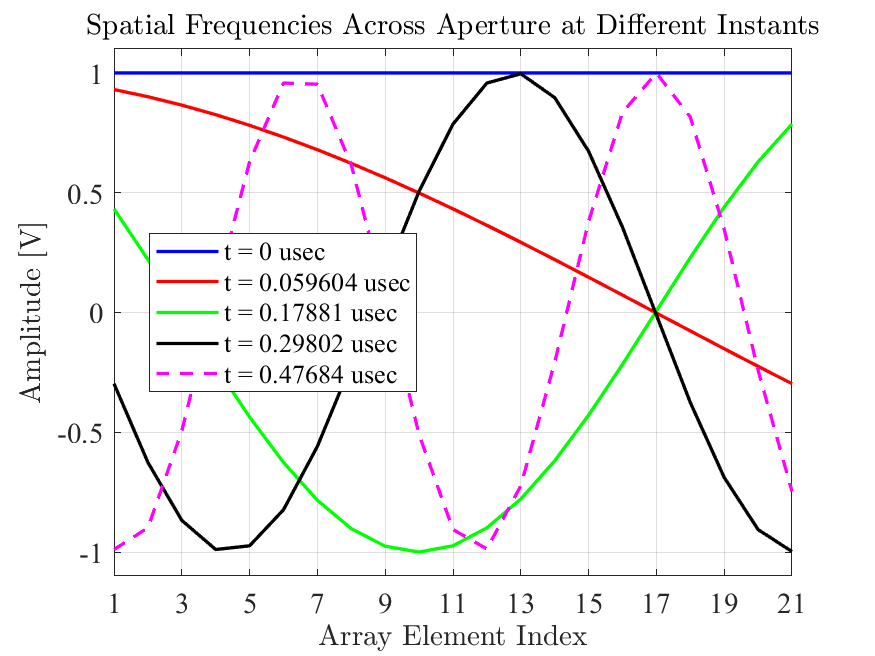}
  \caption{Progression of spatial frequencies across the aperture as a function of time.  In this case, the frequency comb consists of 21 sinusoids from 1 to 5 MHz in steps of 0.2 MHz.}
  \label{fig:fig04}
\end{figure}

\begin{figure}[]
  \centering
  \includegraphics[width=8cm]{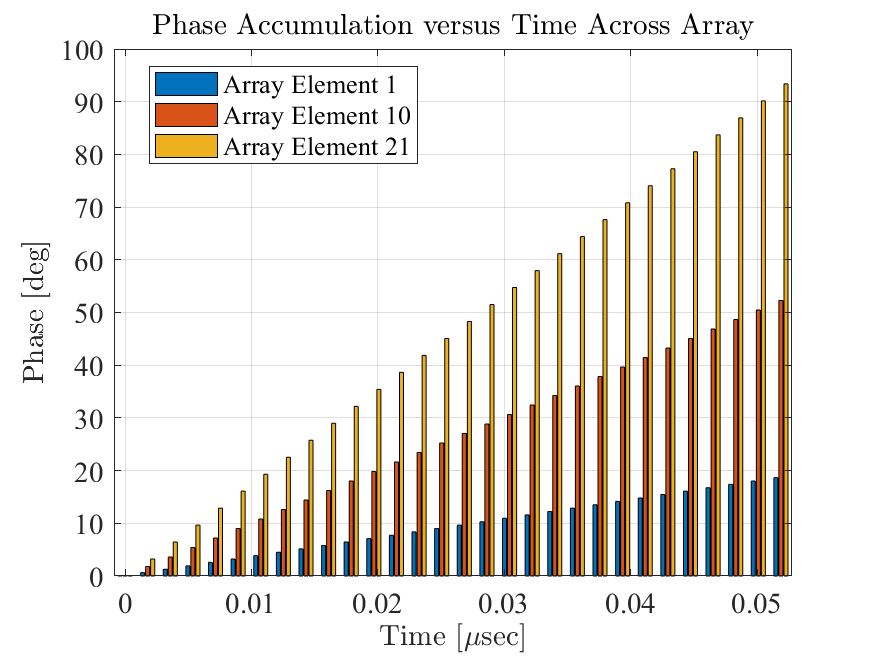}
  \caption{Bar graph shows the array elements matched to higher frequencies accumulate phase faster. 
 As time increases over the duration of the received comb waveform, progressively higher spatial frequencies are generated across the aperture.}
  \label{fig:fig05}
\end{figure}

Since each array element is tuned to a different frequency they will accumulate phase at different rates over the duration of the received waveform.  The beamforming operation sums together the real element outputs directly at the carrier frequency or after a mixing operation removes ${f_0}$ such that the comb is centered at 0 Hz.  At each instant in time a different spatial frequency is created across the array as illustrated in Fig. \ref{fig:fig04}.  As time increases, the high frequency array elements accumulate more phase and the spatial frequency created across the array grows higher as shown in Fig. \ref{fig:fig05}.  At one precise time instant all the phases across the array align to yield a peak output amplitude corresponding to the signal source.

Consider the case of of a linear array with ${N}$ elements corresponding to ${N}$ tones in the frequency comb.  The time axis from 0 to ${T}$ corresponds to the spatial frequencies ${-1 \leq k_{x} \leq 1}$.  The spatial frequency ${k_{x}}$ can be converted to azimuth angle coordinates according to,
\begin{equation}
\label{E:eqn11}
     \tan{AZ} = \frac{k_{x}}{\sqrt{1 - k_{x}^2}}.
\end{equation}
Note the time axis at the output of the beamformer is periodic and peaks will repeat every ${T}$ seconds.  Thus, the maximum unambiguous value of ${T}$ is equal to ${1/{\Delta}f}$ seconds.  

\section{Simulated Results}
\begin{figure}[]
  \centering
  \includegraphics[width=8cm]{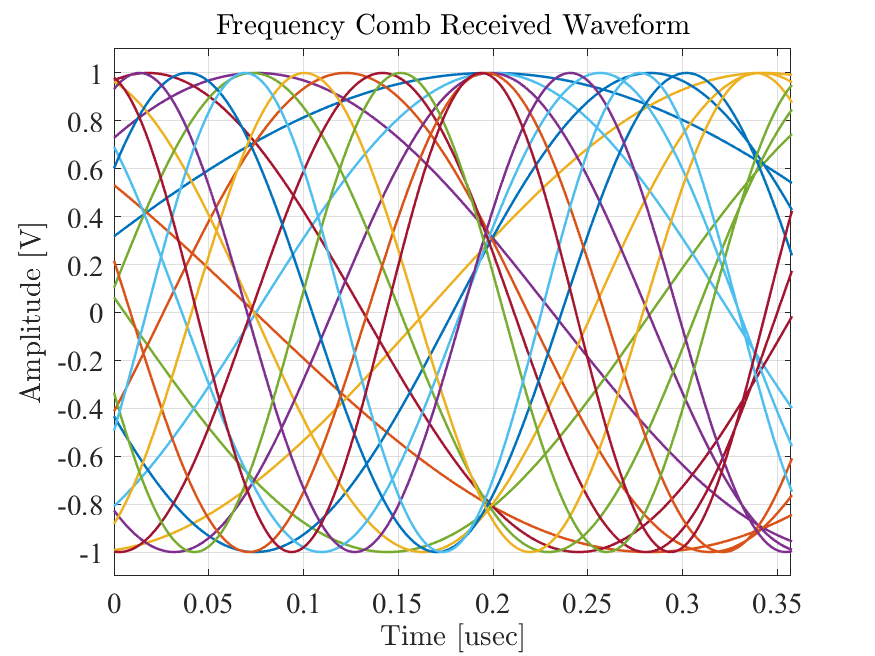}
  \caption{Frequency comb received waveform showing initial phase shift due to propagation delay.}
  \label{fig:fig06}
\end{figure}

In this section we provide simulated results for a linear array along the horizontal x-axis with 21 elements.  Each array element starting from the left edge at ${x=0}$ is matched to a single sinusoid in a frequency comb that varies from 19.001 GHz to 19.005 GHz in steps of 0.2 MHz.  The spacing between array elements is ${\lambda/2}$ at 19.005 GHz and the duration of the comb waveform is 5 ${\mu}$sec.  There is a single signal source of unity amplitude at spatial coordinates ${(x=-6, y=0, z=6)}$ meters where ${z}$ corresponds to the boresight direction and the ${y}$-axis is in the vertical direction.  The ground truth azimuth angle of the signal source is ${-45^{\circ}}$ and its distance from the array origin at ${x=0}$ is 8.4853 meters or 28.3 nsecs.  Recall however that the time axis at the output of the beamformer will not correspond to distance or delay, but rather to k-space or angular coordinates.  Fig. \ref{fig:fig06} illustrates the received comb waveform.  The initial phase of the ${k}$th sinusoid corresponds to the propagation delay between the signal source and the ${k}$th array element at the ${k}$th comb frequency.

\begin{figure}[]
  \centering
  \includegraphics[width=8cm]{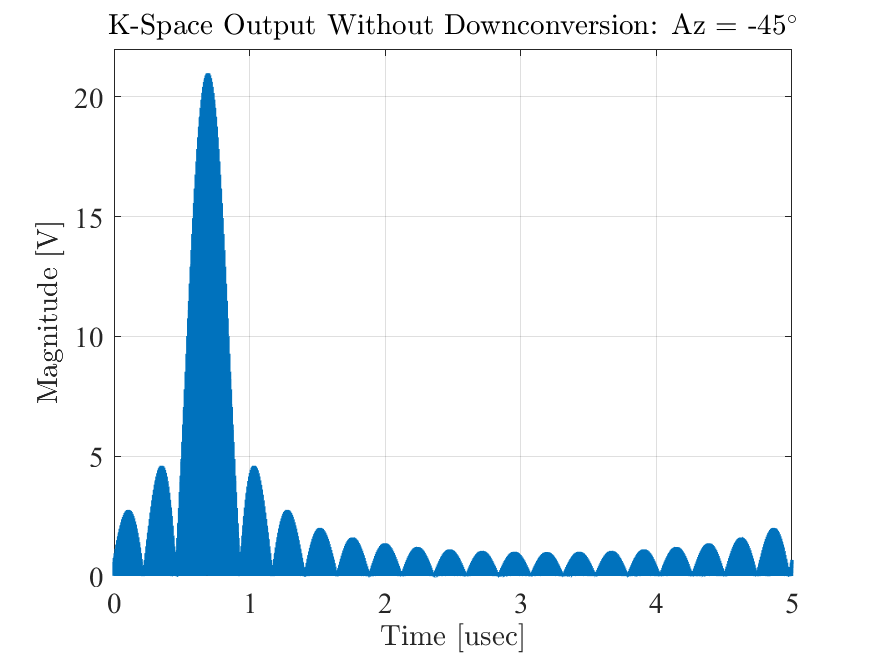}
  \caption{Beamformer k-space output at RF.}
  \label{fig:fig07}
\end{figure}

\begin{figure}[t]
  \centering
  \includegraphics[width=8cm]{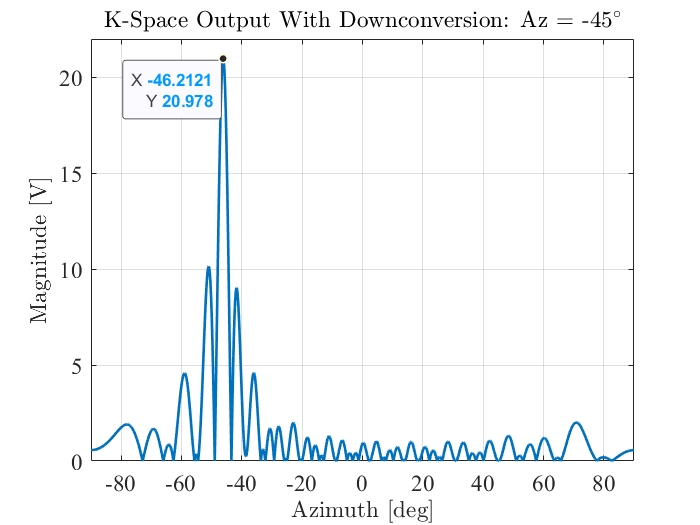}
  \caption{Beamformer k-space output after removing the carrier and converting the time axis to azimuth angle coordinates.}
  \label{fig:fig08}
\end{figure}

Fig. \ref{fig:fig07} illustrates the beamformer output versus time after summing the real RF signals across all the array elements.  Fig. \ref{fig:fig08} illustrates the beamformer output after mixing the RF signals with a local oscillator (LO) signal at 19 GHz and converting the time axis to azimuth angle using (\ref{E:eqn11}).  As can be seen there is a slight error in the estimated angle of the signal source which will be discussed later.

\begin{figure}[]
  \centering
  \includegraphics[width=8cm]{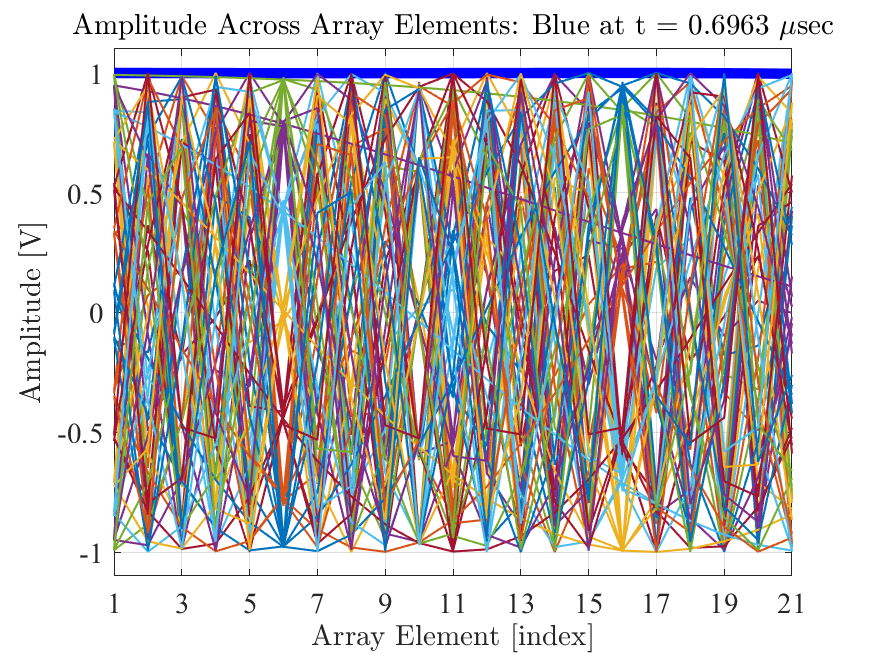}
  \caption{Signal amplitudes across array elements at different time instants.  The blue curve corresponds to ${t=0.6963}$ ${\mu}$sec and yields a peak in the beamformer output.}
  \label{fig:fig09}
\end{figure}

\begin{figure}[]
  \centering
  \includegraphics[width=8cm]{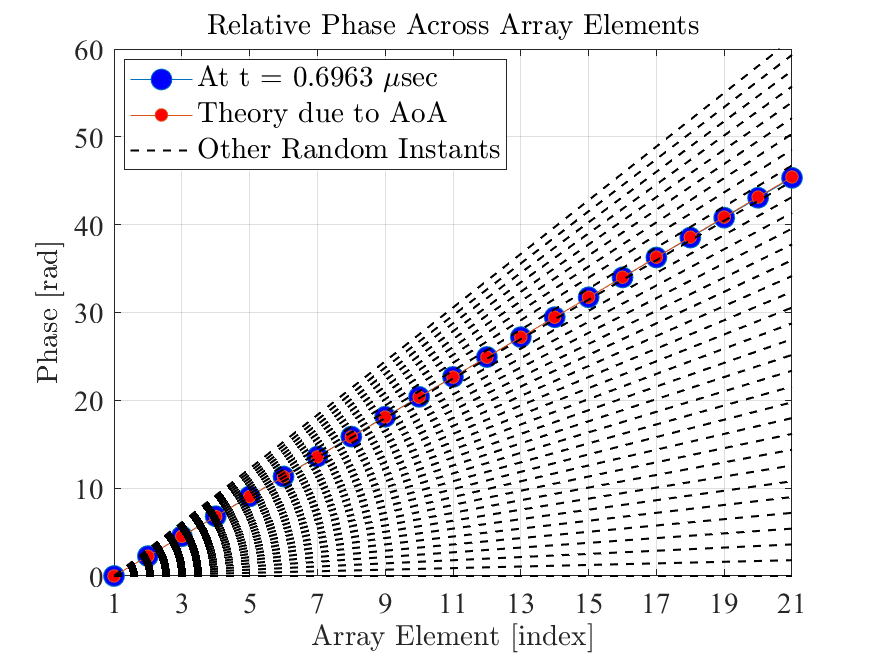}
  \caption{Plot showing relative phase shifts between array elements for different time instants.  At ${t=0.6963}$ ${\mu}$sec the relative phase shifts correspond almost exactly to the theoretical values computed for the AoA of the signal source.}
  \label{fig:fig10}
\end{figure}

Fig. \ref{fig:fig09} illustrates signal amplitudes across the array elements for random time instants.  At the precise time ${t=0.6963}$ ${\mu}$sec when the sinusoidal phases yield maximum amplitude simultaneously across all array elements, the signal peak is created in the beamformer output.
Fig. \ref{fig:fig10} shows the relative phase shift between array elements at different time instants.  Note that at ${t=0.6963}$ ${\mu}$sec the linear phase taper across the aperture matches closely with the theoretical phase shifts expected for the signal's angle of arrival (AoA).

\begin{figure}[]
  \centering
  \includegraphics[width=8cm]{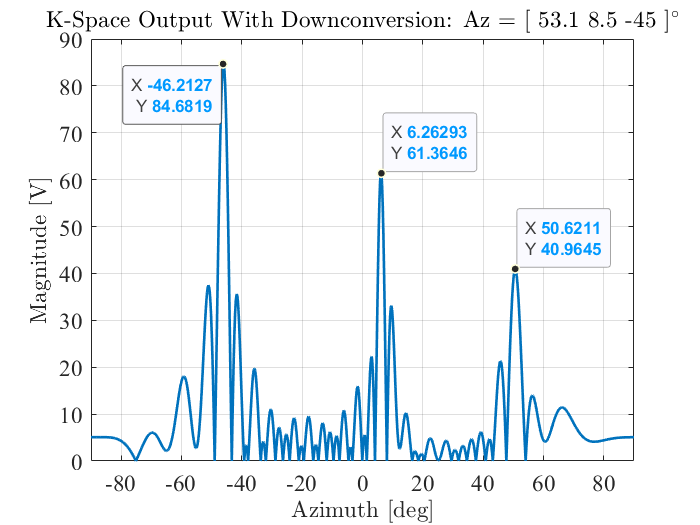}
  \caption{Beamformed output showing 3 signal sources.}
  \label{fig:fig11}
\end{figure}

\begin{figure}[]
  \centering
  \includegraphics[width=8cm]{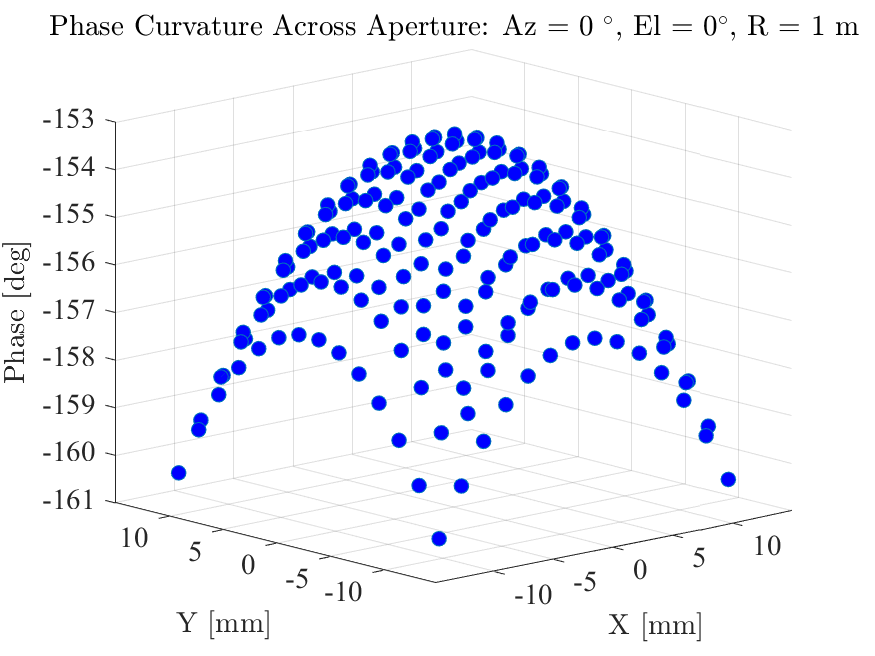}
  \caption{Curvature of phase front across array for point signal source at a distance of 1 m.}
  \label{fig:fig12}
\end{figure}

Fig. \ref{fig:fig11} illustrates the output of the beamformer for 3 signal sources located at azimuth angles ${53.1^{\circ}, 8.5^{\circ}, -45^{\circ}}$ at a distance of 25, 20.2 and 8.5 meters from the array.
As given by (\ref{E:eqn01}), the phase of a plane wave arriving at an array element depends on the distance traveled.  This dependence on distance can impart a curvature to the phase front of the propagating wave, especially at close distances, and induces errors in the angle estimates of a k-space beamformer.  For example, for a signal source at 1 meter, Fig. \ref{fig:fig12} illustrates the phase curvature across an array of 14-by-14 elements spaced ${\lambda/2}$ apart at 19 GHz.  

\section{Conclusion}
This paper proposes a k-space beamforming approach for use with an array of Rydberg quantum sensors.  By transmitting a comb waveform and assigning a unique frequency to each array element, different spatial frequencies are generated across the aperture.  When the element phases align, a strong peak is created in the array output.

\clearpage
\bibliographystyle{IEEEtran}
\bibliography{references}

\end{document}